# Bio Inspired Approach to Secure Routing in MANETs

*V. Venkata Ramana, **Dr. A. Rama Mohan Reddy, and ***Dr. K. Chandra Sekaran

*SSITS, Rayachoty, Kadappa, India. `ramanajntusvu@gmail.com`
** SVUCE, Tirupati, India. `ramamohansvu@gmail.com`
*** NITK Surathkal, India. `kchnitk@gmail.com`

*Abstract*

*In this paper, the author explore the challenges with respect to the security aspect in MANETs and propose a new approach which makes use of a bio-inspired methodology. This paper elaborates various attacks which can be perpetrated on MANETs and current solutions to the aforementioned problems, and then it describes a Bio-Inspired Method which could be a possible solution to security issues in MANETs.*

*Keywords:*

*MANET, Bio-inspired approach, Agents and Secure Routing*

## 1. INTRODUCTION

A Mobile Ad-Hoc Network (MANET) is a collection of wireless nodes that can be connected to each other dynamically anytime and anywhere without the requirement of the existence of architecture to support the network. It is a self organizing system made of mobile nodes that communicate with each other using wireless link with no central administration such as base stations or access points. [1] Nodes in a MANET act as both hosts and routers to forward packets to each other. [2] The nodes that are within the radio range of each other communicate directly while others use intermediate nodes as relay points. These networks have gained ample interest in recent times due to its various advantages as compared to the networks that require a basic infrastructure to work.

The promise held by the application of wireless ad hoc networks is immense. It ranges across the horizon and the number of real world problems that could be solved with the application of Mobile Ad hoc Networks is growing by the day. A network of this kind is well suited for highly critical applications like disaster management, emergency relief, military operations, mining activities and terrorism response where no pre deployed infrastructure for communication exists. [2] For example, in the case of an earthquake, ad hoc networks could be used for communication when conventional communication networks could be damaged. [3]

## 2. SECURE ROUTING IN MANETS

The very nature of Mobile Ad Hoc Networks brings with it some specific security challenges. The absence of centralized access control, secure boundaries (as mobile nodes are free to leave or join a network) and restricted resources make them vulnerable to various active and passive attacks. They are more susceptible to information and physical security threats than wired networks or infrastructure based wireless networks.

The security concerns involve anonymity, non-repudiation, access control, trust issues, authenticity, confidentiality and integrity. [1] Attacks on a wireless ad hoc network can come from any direction and target any node. Damage to a node can leak information, contaminate message and impersonate nodes. Nodes have adequate physical protection and are receptive to being compromised as mobile nodes are autonomous and are free to roam independently. [3] Moreover, the decision making for routing is taken by independent nodes in a co-operative fashion without the interference of a centralized authority. Hence it is susceptible to attacks that target the co-operative algorithm. [2] Attacks on the routing protocol could be either internally or externally generated. External attack could be through the injection of faulty routing information, replaying old information or distorting routing information which causes a traffic overload resulting in retransmission and inefficient routing. The internal threat could be from compromised nodes which might misuse routing information and induce service failures. Since attacker is already part of the network, internal attacks are more severe and hard to detect than external attacks.

A closed group of nodes could be secured using certificates. Distributed security schemes mainly rely on threshold cryptography. In an ad hoc environment relying on gateway nodes connecting to e.g. the Internet more centralized security schemes can be applied as well. Many protocols using some sort of cryptographic certificates leave the questions concerning certificate distribution, management, and especially revocation untouched.

The absence of infrastructure and the consequent absence of authorization facilities impede the usual practice of establishing a line of defense, separating nodes into trusted and non-trusted. Such a distinction would have been based on a security policy, the possession of the necessary credentials and the ability for nodes to validate them. In the MANET context, there may be no ground for an a priori classification, since all nodes are required to cooperate in supporting the network operation, while no prior security association can be assumed for all the network nodes.

Additionally, in MANET freely roaming nodes form transient associations with their neighbors, join and leave MANET sub-domains independently and without notice. Thus it may be difficult in most cases to have a clear picture of the ad hoc network membership. Consequently, especially in the case of a large-size network, no form of established trust relationships among the majority of nodes could be assumed. In such an environment, there is no guarantee that a path between two nodes would be free of malicious nodes, which would not comply with the employed protocol and attempt to harm the network operation. [4]

Network layer attacks could be impersonation (masquerading or spoofing), modification, fabrication and replay of packets. Specifically, fabrication attacks where an intruder generates false routing information in order to disturb network operation or to consume other node resources. [5]

Fabrication attacks could be:

i. Resource Consumption attack - Here a malicious node deliberately consumes the resources (battery power, CPU cycles, bandwidth) of other nodes in the network. It can flood the network with routing messages. The main goal of this attack is to reduce the network capabilities as much as possible. It also results in CPU cycle and battery wastage when the nodes update their routing tables constantly. A more subtle variation of this attack is when a node behaves selfishly and gives false information so as to not forward packets in order to save its own battery power. [5]

ii. Black Hole Attack - In this attack a malicious node advertises itself as having the shortest path (or most stable path) to all nodes in the environment by sending fake route reply exploiting the ad hoc routing protocol such as AODV. [2] It can be used as a denial-of-service attack where it will drop the packets later. Co-operative black hole attack is when several malicious nodes co-operate with each other for the attack making the detection of such attacks more difficult.

iii. Grey hole attack - This attack is similar to black hole attack except it drops the packets that are intercepted only with a certain probability making it more surreptitious than black hole attack. With this selective dropping the intruder tries to hide the attack by not denying all the network services. This attack is harder to identify because the reduction of network capabilities due to the attack could be mistake for normal instability from wireless connections. [5]

iv. Worm hole attack - A malicious node tunnels a received packet from one point in the network to another and then replays them into the network from that point. [6]

v. Routing table overflow - The attacker attempts to create routes to nonexistent nodes. The goal is to create enough routes to prevent new routes from being created or to overwhelm the protocol implementation. [7]

vi. Sleep Deprivation - An attacker can attempt to consume batteries (which is important in MANET) by requesting routes or by forwarding unnecessary packets to the node. [7] This sort of attack is a serious security issue since it can act as a denial-of-service attack or has the potential to be the first step in Man-in-the-middle attack. Its consequences can vary from having one node disconnected from the entire network to all the nodes in the network relaying on the malicious node being totally disconnected to each other.

## 3. DIFFERENT APPROACHES TO SECURITY IN MANETS

i. Secure Routing Protocol [8] - It only counters the malicious behavior that targets the discovery of topological information. SRP provides the correct routing information about a pair of nodes provided they have prior security association. SRP uses minimal trust to avoid black hole attack. Grey hole attack is avoided by packet leashes.

ii. Secure Efficient Ad hoc Distance Vector Routing Protocol [9] - It uses efficient one way hash functions to guard against Denial of Service attacks. The hash chain elements are used to authenticate the metric and the sequence number of a routing table. The

153

authentication could either be Message authentication codes or some broadcast authentication mechanism. It is robust against multiple uncoordinated attacks, but fails against the wormhole attack.

iii. Mitigating Routing Misbehavior [10]

- Watchdog - This solution is based on the idea that the node is capable of listening to the information of how the neighboring node (which claimed to have the shortest path to the destination) is routing the packet. This can be done by switching into promiscuous mode after sending the package so as to listen to the transmissions of its neighbors in order to verify if the router actually transmitted the package.
- Pathrater - In this system each node uses the information received from systems like watchdog to rate its neighbours according to its trustworthiness. This information will help the node to forward packages to trusted nodes in the future.

a) Ariadne (On demand reactive routing) [11] - It uses symmetric source routing and prevents DoS and Route tampering. The protocol assumes nodes will only trust themselves and the destination node for path authentication. It can protect against black hole attack as the paths are authenticated only by the destination and hence malicious nodes are excluded from the path.

b) Routeguard [12] - This method classifies its neighboring nodes as Member, Unstable, Suspect, or Malicious from the data obtained from Watchdog and Pathrater. Each class or tag implies a different trust level ranging from trusted (Member) to Untrusted (Malicious).

However all these systems are based on the assumption that every node transmits with an Omni-directional antenna. It has been shown that with a directional antenna to a legitimate node, the malicious node can override the watchdog detection system and perform a Black Hole attack.

i. Another proposed solution for the black hole attack is to find more than one route to the destination. (Redundant routes). [13] The source node then unicasts a ping request to the destination using the three routes. The source will process the acknowledgements received from these packets in order to figure out the route which is not safe. The sender node needs to verify the authenticity of the node that initiates the RREP packet by utilizing the network redundancy. This solution can guarantee to find a safe route to the destination but the main drawback is the time delay. Besides, if there are no shared nodes or hops between the routes the packets will never be sent.

ii. Patwardhan secure routing and intrusion detection system [14] - This presented a proof of concept where a secure routing protocol was implemented using public key encryption, intrusion detection, and a reaction system. The system implements a secured routing protocol adding public key signatures to verify the ownership of the messages. By addition, it has an intrusion detection system where each node monitors its neighbors in promiscuous mode listening to their routing activity. When a node claiming to be a router, is detected misbehaving, the detection system marks the node as malicious node and the reaction system isolates the node from the ad-hoc network.

iii. A solution has also been proposed for multiple black hole nodes acting in cooperation. [15] It involves sending additional 2 bits information from the nodes responding to the RREQ (Route Request). Each node will also maintain an additional Data Routing Information (DRI) table. However the performance of the proposed solution has not been analyzed using simulations. The solution is mainly theoretical.

iv. AODV-SEC [16] - It is a secure routing protocol for MANETs based on AODV (Ad hock On Demand Distance Vector Routing) called AODV-SEC. It uses certificates and a public key infrastructure as trust anchor. In addition it presents the need for a new certificate type for secure routing in MANETs called mCert. The AODV-SEC protocol tries to secure all possible aspects of the route discovery process. This includes the authentication of the two end nodes as well as the intermediate nodes. Further, it excludes not trusted nodes from the discovered routes. The length of the discovered route is protected in a way that intermediate nodes cannot advertise a potentially shorter route than actually exists.

## 4. DESIGN FOR SECURE ROUTING IN MANETS USING BIO-INSPIRED APPROACH

Bio-inspired approach is an emerging field in problem solving techniques. Bio-inspired algorithms have the unique feature of being bottom-up, adaptable and flexible, thus providing elegant solutions to engineering problems that are constrained by rigid limitations that traditional approaches pose. Several bio-inspired algorithms work as highly decentralized systems consisting of several components. MANETs too are almost completely decentralized systems and consist of mobile nodes that are components of the system. Thus, bio-inspired techniques can be accurately applied to MANETs to solve various issues in the design and working of MANETs.

The immune system has features that enable it to detect, isolate and destroy infections as well as assimilate information about recent infections in the body. These features are highly desirable in a security system for MANETs. To replicate the behavior of the immune system in a MANET, the following characteristics have to be present in the system [19]

- Ability to recognize a malicious node
- The ability to prevent, detect and eventually eliminate dangerous foreign activities or infections.
- Retain previous infections in memory
- Mechanism to identify new forms of infection.
- Methods to autonomously come up with responses to handle the infection.
- Prevent the system from self attacks.
- Ability to find an alternate routing path.
- Ability to achieve network security while keeping additional traffic to minimum.

To achieve the above characteristics implementation of the following steps is required:[19]

i. Agents: Different nodes should generate different types of agents (similar to the agents of immune system). Each agent is capable of handling a different type of malicious behavior (infection in immune system).
ii. Self Test: The agents generated by a node are passed through a self test to detect anomalies in the agent. This self test is conducted periodically on the agent to detect

155

           foreign code injection. If the agent does not pass the self test it is declared corrupt an eliminated.(Negative selection in Immune system)
iii. Broadcasting: Each node broadcasts its agents to some other nodes in the system. Each node will have a copy of other nodes agent and thus creating a security database at each node. If any malicious behavior (foreign event) is detected, the agent notifies the other nodes in the system by broadcasting the event. This enables a faster response to the event in the future.
iv. Agent life cycle: Each agent has a Time to Live (TTL) feature to avoid the flooding of the network with agents. The timer inherent in each agents controls the lifetime of the agent. The TTL is different for different types of agents depending on the infection it is meant to tackle.
v. Approximate matching: After detecting a foreign event, the agent evaluates it's affinity towards the event depending on its ability to solve that particular infection. This mechanism is called approximate matching. Hamming distance and bit patterns can be used as matching schemes.
vi. Handling malicious nodes: Depending on the degree of affinity, elimination of the foreign event is done. If the matching is not proficient, the event is broadcasted to all other agents and the best agent to tackle the problem is determined.
vii. Source node identification: The source node of the malicious event is identified. If the same source node, has been involved in previous infections the source node is declared malicious (un-trusted) and is isolated from the system.

### 4.1.Design Details

Agent - An agent is a packet containing a specially designed program that performs the following functions:

a) Detects any sort of malicious behavior in nodes
b) Tackles a specific problem or infection in a node

Architecture of the agent: The agent is generated by a node in the system. This node is called the source node of the agent. The agent contains a program binary/code which is responsible for detection and correction of malicious nodes. The agent also contains details of its affinity to different infections. The agent maintains its Time-To-Live status and its Test Key which will be read by receiving nodes for performing the self test function.

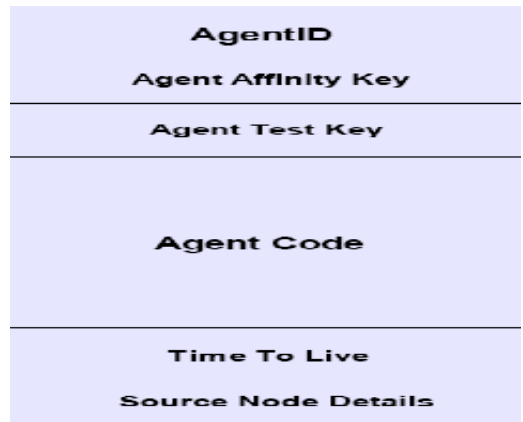

The agent will have the following structure ( pseudo code format)

```
/*
* Structure to define the agent ID
* Contains Source ID, type of the agent based on the infection it tackles
* and a count field which uniquely identifies the agent within the source node
*/
typedef struct agentid
{
 long int source;
 long int agenttype;
 long int count;
} aId;

class agent{
  aId id; //Agent id
  long int affkey; // A key defining the agent's affinity towards a particular infection
  long int testkey; // A key required for the self test
  long int ttl; //Time to live defines life time of the agent
  long int srcnode;   //Unique identifier of the source node
};
```

Functioning of the agent can be understood as per the following :

- At the Source Node - The source node of the agent first tests its neighboring nodes by running the agent code. The agent code sniffs the behavior of the neighboring nodes by putting the source node in promiscuous mode. Based on listening to the node's neighbors' routing activity, it declares the neighbors as "clean" or "probably infected" and updates this information in the node's security database. The node then sends the agent or copies of the agent to the nodes that have been declared "clean" in its database.

- At Other Nodes - Once an agent arrives at a node, the code in the agent is extracted by the node. The node then runs the code in its system and detects the behavior of its neighboring nodes. The result of the sniffing is updated in the node's security database.

Detecting Malicious Behavior - Malicious behavior in a node is detected using several parameters such as number of dropped packets, rate of transmission of packets etc depending on the specific malicious behavior that the agent is designed to detect and tackle.

Reacting to Malicious Behavior - If an agent designed to detect a particular malicious behavior, finds a node exhibiting that behavior, it declares the node as "probably infected". The agent then broadcasts this information in the network. All the nodes that receive the broadcast packet will update their security database to include the malicious behavior of that particular node.

If broadcast information of a particular node that has already been declared as probably infected, arrives at a node, then the node is declared as Infected. This particular node is eliminated from the Mobile ad hoc network as the neighboring nodes stop sending packets to the node that is declared malicious.

Self test - The purpose of the self test phase is to ensure that the agent packet itself has not been corrupted or injected with foreign code. Self test can be conducted on an agent through the technique mentioned below.

We will use a technique where a large arbitrary stream of data is converted into a unique short bit string data. An agent tackling a particular malicious behavior will have a unique fingerprint. A global table is maintained which maps all the agents to its unique fingerprint. This table also contains details of all the malicious behaviors that can be detected and tackled by the malicious node.

Every node that receives an agent will run a test on that particular agent to ensure that the agent has not been compromised. This test will be based on cryptographic protocols. Only if the agent passes the test will the node run the agent code in its system.

Our approach is as follows: The "global table" can have agent id, encrypted H(Hash ) and source node name/public key. We can protect the global table such that entries can only be added and not modified. It will ensure that no one tampers with existing agent details

Agent creator  node : generates Public Key, private key. The public key cab be kept in a global table. Applies hash function f on C(Agent Code), generates F. Encrypts F with private key = H. Keeps H also in the global table.

Receiver: Decrypts H (from the global table) using the public key to get G. Apply same hash function f on the code and get F. Compare F & G.

Time-To-Live - Each agent has a Time to Live parameter attached to it. When the count reaches 0, the agent self-destructs after broadcasting all the relevant information to all nodes in the network so as to retain all important information such as infection-history. This is done by the node where the agent currently resides or the node that is the next target destination of the agent. The destruction is done by terminating the propagation of the agent to the next "clean" node and by killing the agent code that is running in the current "clean" node.

Approximate Matching - When an infection is detected by an agent, it decides its ability of tackling that particular malicious behavior based on hamming distance. The hamming distance is calculated between the agent affinity key (stored in the agent) and the infection key (stored in the security database).

This gives rise to the following two situations:

  i.  If the hamming distance is the least then the agent has an affinity to the infection and tries to tackle the infection.
  ii. If the affinity to the infection is less (hamming distance is high), the agent broadcasts the infection key and the IP of the source malicious node to the network. Based on this broadcast all the nodes in the network updates their security databases. Agents at all nodes receive this information and the agent that is physically closest to the infected node travels to the node and attempts to tackle the infection there.

158

Advantages of the Approach are:

 i. Scalability - The Security Database is a distributed table that stores the most updated status of all the nodes and agents in the system. This table is built using a local Security Database at each node and a look up system to achieve consolidation and visibility of all information. This will solve the problem of scalability of security information in the nodes.

 ii. Phase-by-phase implementation - The implementation of this system can be carried out phase by phase. The complete infrastructure of a MANET need not be changed for this intrusion system to be implemented. This security system can run even if only few nodes of the MANET have the bio inspired methodology incorporated in them.

 iii. Zero-trust policy - All nodes start out with the status of being "probably infected" except the initial node that begins the MANET. Thus, each node is tested thoroughly before being accepted into the MANET.

 iv. Expansion - Methods for countering newly discovered infections can easily be incorporated into the security system by creating new types of agents for the new infections. The underlying infrastructure of the security system is unaffected by the number or type of infections that it is expected to handle. It is thus extremely flexible.

## 5. CONCLUSION

This paper gives a comprehensive overview of the security problems of MANETs. The current research scenario in this area is still in its nascent stages. There is indeed immense potential to implement a Bio Inspired approach to the security problem of MANETs. The immune system provides a good parallel to the security needs of a MANET, and thus we can draw inspiration from the human immune system to develop a secure routing system for MANETs.

## REFERNCES

## Authors

V. Venkata Ramana obtained his Master's degree in Computer Science Engineering form JNTUH, University, pursuing Ph.D. degree in JNTUA, University and at present working as an Associate Professor in Department of Computer Science Engineering, SSITS, Rayachoti, Kadapa(Dist), A.P. His areas of interest include Computer Networks, Mobile Adhoc Networks, Bio-Inspired Networks and other latest trends in  technology. He has more than 09 years of experience in teaching and research in the area of Computer Science and Engineering.

Dr. A Rama Mohan Reddy working as Professor and Head in the Dept. of  Computer Science and Engineering, Sri Venkateswara College of Engineering(SVUCE), Sri Venkateswara University., Tirupati, Chittor(Dist.), AP-INDIA.He Completed his M.Tech Computer Science from NIT, Warangal and completed his  Ph.D in Software Architecture from SV University, Tirupati. His areas of interest include Software Architecture, Data Mining, Computer Networks, Mobile Adhoc Networks, and other latest trends in  technology. He has more than 28 years of experience in teaching and research in the area of Computer Science and Engineering.

Dr. K.Chandra Sekaran is working as Professor in the Department of Computer Science & Engineering at National Institute of Technology Karnataka, Surathkal, India. He has 25 years of teaching and research experience in the field of computing and informatics. He has more than 100 papers published in reputed journals and International proceedings.